\documentclass[review,sort&compress]{elsarticle}

\usepackage{lineno,hyperref}
\usepackage{textcomp}
\usepackage{setspace}
\usepackage{color}

\journal{Colloids and surfaces A: Physicochemical and engineering aspects}

\begin{document}

\begin{frontmatter}

\title{
Spreading of Aqueous Droplets with Common and Superspreading Surfactants. A Molecular Dynamics Study
%\tnouperspreadteref{mywiing andrfactants common th stsuitlenote}
}


%\author[panos]{Panagiotis E. Theodorakis\corref{cor1}}
\author[panos]{Panagiotis E. Theodorakis}
%\cortext[cor1]{panos@ifpan.edu.pl}

%\author[edward]{Edward R. Smith\corref{cor2}}
\author[edward]{Edward R. Smith}
%\cortext[cor2]{edward.smith@brunel.ac.uk}

%\author[chemeng]{Erich M\"uller\corref{cor3}}
\author[chemeng]{Erich A. M\"uller}
%\cortext[cor3]{e.muller@imperial.ac.uk}

%\author[maths]{Richard V. Craster\corref{cor4}}
%\author[maths]{Richard V. Craster}
%\cortext[cor4]{r.craster@imperial.ac.uk}

%\author[chemeng]{Omar K. Matar\corref{cor5}}
%\author[chemeng]{Omar K. Matar}
%\cortext[cor5]{o.matar@imperial.ac.uk}

\address[panos]{Institute of Physics, Polish Academy of Sciences, Al.\ Lotnik\'ow 32/46, 02-668 Warsaw, Poland}
\address[edward]{Department of Mechanical and Aerospace Engineering, Brunel University London, Uxbridge, Middlesex UB8 3PH, United Kingdom}
\address[chemeng]{Department of Chemical Engineering, Imperial College London, Exhibition Road, South Kensington, London SW7 2AZ, United Kingdom}
%\address[maths]{Department of Mathematics, Imperial College London, Exhibition Road, South Kensington, London SW7 2AZ, United Kingdom}

%\singlespacing

\begin{abstract}
The surfactant-driven spreading of droplets is an essential process in many applications ranging from 
coating flow technology to enhanced oil recovery. Despite the significant advancement
in describing spreading processes in surfactant-laden droplets, including the exciting
phenomena of superspreading, many features of the underlying mechanisms require further understanding.
Here, we have carried out molecular dynamics simulations of a coarse-grained model with 
force-field obtained from the Statistical Associating Fluid Theory to study droplets laden with common
and superspreading surfactants. We have confirmed
the important elements of the superspreading mechanism, \textit{i.e.} the adsorption of
surfactant at the contact line (CL) and the fast replenishment of surfactant from the bulk. 
Through a detailed comparison of a range of droplets with different surfactants, 
our analysis has indicated that the ability of surfactant to adsorb at the interfaces
is the key feature of the superspreading mechanism. To this end, surfactants that tend to form aggregates
and have a strong hydrophobic attraction in the aggregated cores
prevent the fast replenishment of the interfaces, resulting in reduced spreading ability. 
We also show that various surfactant properties, such as diffusion and architecture, play a
secondary role in the spreading process. Moreover, we discuss various drop properties such as
the height, contact angle, and surfactant surface concentration highlighting
differences between superspreading and common surfactants.
We anticipate that our study will provide further insight
for applications requiring the control of wetting.
\end{abstract}


\end{frontmatter}

%\linenumbers

\section*{Introduction}
%\subsection*{Theory}
Superspreading of surfactant-laden aqueous droplets is an exciting phenomenon, 
which has received a great deal of attention over the last six 
decades \cite{Hill1998, Nikolov2011, Schwarz1964,Venzmer2011,Theodorakis2014,Sankaran2019}. 
It refers to the unexpectedly rapid spreading of aqueous droplets on hydrophobic substrates, 
due to the presence of surfactant molecules known as superspreaders \cite{Ananthapadmanabhan1990,Rafai2002}. 
This phenomenon is of fundamental importance for diverse applications,
such as coating technology, drug and herbicides delivery, and enhanced oil
recovery \cite{Nikolov2011,Bonn2009,Craster2009,Matar2009,Rosen2012}.
Although the first reports of superspreading date back to over
50 years ago \cite{Schwarz1964}, this phenomenon still attracts considerable
attention from both theory and
experiment \cite{Venzmer2011,Theodorakis2014,Holder2014,Theodorakis2015a,Theodorakis2015b,Badra2018,Wei2018,Kovalchuk2014,Kovalchuk2016a,Kovalchuk2016b}. 
While experimental \cite{Kovalchuk2014,Kovalchuk2016a,Kovalchuk2016b, Lee2009}
and theoretical \cite{Theodorakis2015a,Theodorakis2015b,Holder2014,Holder2015,Badra2018,Wei2018}  
studies have discussed possible mechanisms of the superspreading for 
surfactant-laden droplets, certain aspects of this phenomenon require further discussion. This
includes the distribution of surfactant molecules within the droplet and the role of surfactant
aggregation and diffusion in the spreading process. Moreover, simulation studies have thus far only
considered a limited selection of superspreading and common surfactants and a broader selection of 
surfactants would provide more information towards identifying similarities and differences between 
surfactant behaviour. 

The study of spreading phenomena by computer simulation is well justified, given the availability of
reliable all-atom \cite{Hautman1991,Mar1994,Fan1995,Lane2008}
and coarse-grained (CG) models  \cite{Saville1977,Sikkenk1988,Nijmeijer1989,Nijmeijer1990,Nieminen1994a,Nieminen1994b,Ruijter1999,Blake1999a,Blake1999b,Voue1999,Voue2000a,Voue2000b,DeConinck2001,Lundgren2002,Lundgren2003,Heine2003,Werder2003,Marrink2007}
that enable the faithful simulation of these systems. 
Indeed, simulations of aqueous solutions with surfactants \cite{Tomassone2001a,Tomassone2001b,Shinoda2007,Shinoda2008,Herdes2015,Lobanova2016,Lobanova_PhD,Holder2014,Holder2015}
have established the connection between the behaviour of surfactants in the bulk and
spreading \cite{Shen2005,Kim2006,Halverson2009,Sergi2012},
while the superspreading mechanism and the main characteristics of superspreading
surfactants have been the focus of recent studies  \cite{Theodorakis2014,Theodorakis2015a,Theodorakis2015b,Badra2018,Wei2018,Holder2015,Holder2014,Shen2005,Sergi2012,Halverson2009,Smith2018}. 
Moreover, experiments  have elucidated a number of factors that aid or suppress spreading, 
such as the rate of evaporation \cite{Semenov2013}, humidity \cite{Ivanova2012}, pH \cite{Radulovic2009a}, 
surfactant structure and concentration \cite{Ivanova2010,Svitova1998},
surfactant aging effects \cite{Radulovic2010}, surfactant mixtures \cite{Rosen2001,Wu2002},
substrate hydrophobicity \cite{Hill1998,Ivanova2012,Radulovic2009b}, and temperature \cite{Ivanova2010,Ivanova2011}.
Despite numerous experimental and numerical studies on 
the superspreading of surfactant-laden droplets, the study of the
superspreading mechanism requires access to molecular-level information
of the system, which is not accessible to experiment and continuum simulation.
Therefore, we employ
here large-scale Molecular Dynamics (MD) simulation of
a CG model to study the spreading process by using different common and superspreading surfactants (Fig.~\ref{fig0}). 
We perform analysis of various properties highlighting differences and similarities 
in the spreading process for these surfactants.
Our study also highlights the importance of the 
aggregation tendency of surfactants, 
which affects the adsorption efficiency
of surfactants at the interfaces and is an important component of
the superspreading mechanism \cite{Theodorakis2015b,Theodorakis2015a}.
Moreover, properties such as surfactant diffusion and architecture, 
among others, seem to play a lesser role in superspreading.

\section*{Model and methods}
\begin{figure}
\centering
\includegraphics[scale=0.40]{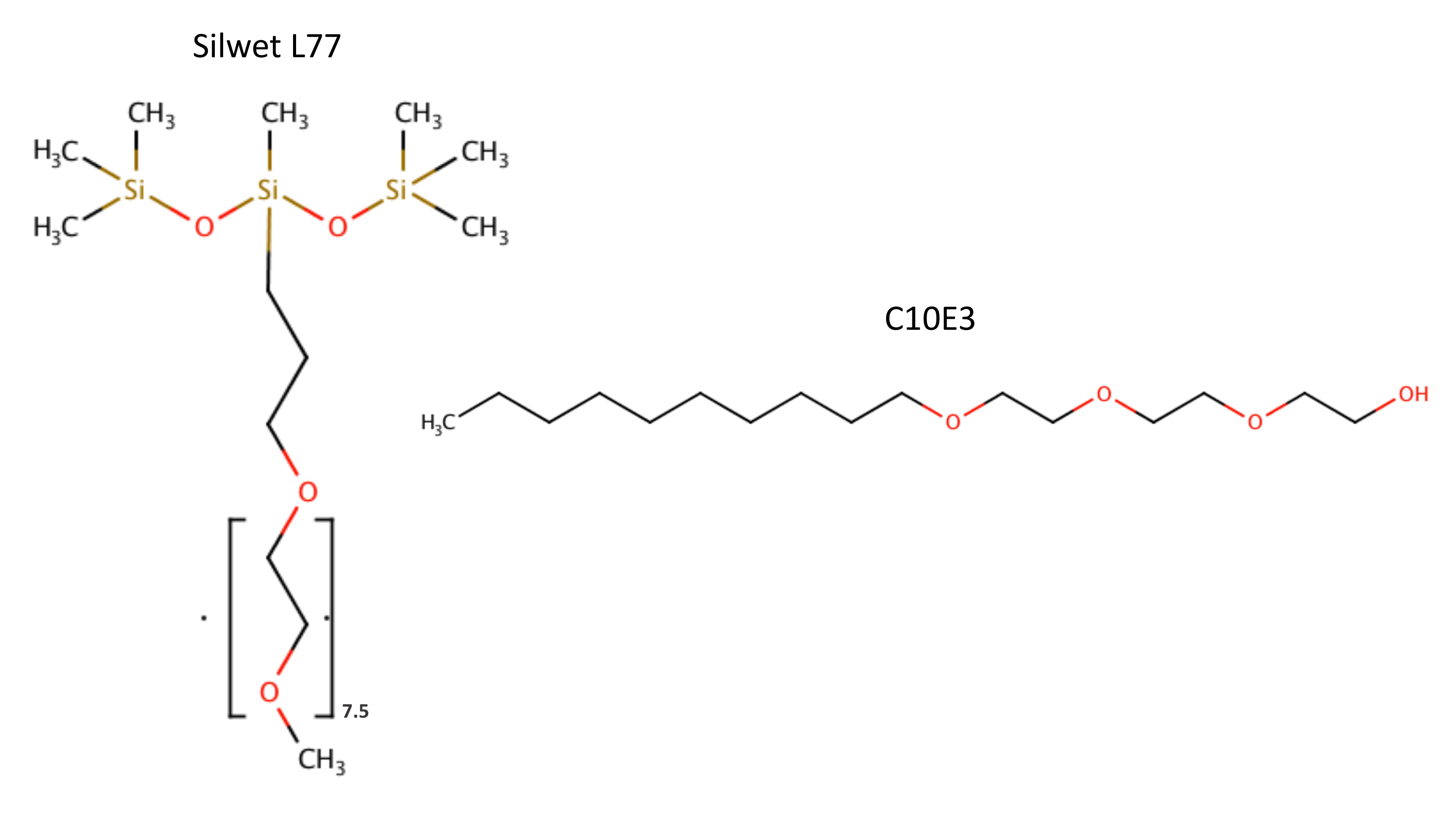}
 \caption{\label{fig0}  Structure of the superspreading Silwet L77 and and 
 common C10E3 surfactants. The T3E3 surfactant has the same structure as the Silwet L77 surfactant, but it
 only consists of three ethoxyl groups. Similarly the number of alkyl and ethoxyl group are
 varied accordingly in the case of common surfactants considered in this study, such as
 C10E8, C12E5, and C12E6.}
\end{figure}

In this study, we carried out MD simulations of a CG model to study systems of 
aqueous droplets laden with either common or superspreading surfactants. 
We have considered a widely used superspreader (Silwet L77) and 
a common surfactant (C10E8) as well as different cases by varying the legnth of the hydrophilic
(common and superspreading surfactants) and the hydrophobic parts (common surfactants) 
of the surfactants (Fig.~\ref{fig0}).
Our CG model stems from the SAFT-$\gamma$ molecular-based equation of state (EoS), which
analytically describes thermophysical data \cite{Papaioannou2014,Lafitte2013}. This
model has proven to be particularly suitable to capture the superspreading of 
surfactant-laden aqueous droplets \cite{Theodorakis2015b,Theodorakis2015a} and bulk properties of
these systems \cite{Theodorakis2015a,Lobanova_PhD,Lobanova2015,Lobanova2016,Rahman2018,Morgado2016}.
The EoS offers an accurate fit for force-field parameters, due to 
the close match between the theory and the underlying Hamiltonian of the system. Hence,
it is able to reproduce macroscopically observed thermophysical properties and describe
accurately fluid--fluid and fluid--solid
interactions \cite{Herdes2015,Lobanova_PhD,Lobanova2015,Lobanova2016,Muller2014,Avendano2011,Avendano2013,Jimenez2017,Herdes2017}.
In fact, the SAFT approach derives robust and transferable
potentials of effective beads that represent groups of atoms or even whole molecules
(\textit{e.g.} water) with the approach being capable of describing heterogeneous chain
fluids \cite{Avendano2013}. Moreover, the interaction parameters are traced to macroscopic properties of
the original segments of the associated pure components \cite{Muller2014}.
Here, an effective bead `W' represents two water molecules ($H_2O$) with mass
$m_W=0.8179m$ \cite{Lobanova2015}, where $m$ is the reduced unit of mass corresponding to 44.052~amu. 
Effective beads `M'
represent  a $(CH_3)_3-Si-O_{1/2}$ chemical group with mass, $m_M=1.8588m$, effective
beads `D'  correspond to $O_{1/2}-(CH_3)_2-Si-O_{1/2}$ groups with mass, $m_D=1.6833m$,
`EO' to $-CH_2-O-CH_2$ (ether) chemical groups with mass, $m_{EO}=m$, and
$-CH_2-CH_2-CH_2-$ (alkane, CM) groups with mass, $m_{CM}=0.9552$. We make no distinction between 
terminal methyl groups and the $CH_2$ groups \cite{Lobanova_PhD}. The 
chemical structures of the common and superspreading surfactants 
considered in our study are illustrated in Fig.~\ref{fig0}.

Effective beads interact via the Mie potential, which is described by the following
relation:
\begin{equation}
\label{Mie}
U(r_{ij})=C\varepsilon_{ij} \left[ \left( \frac{\sigma_{ij}}{r_{ij}}\right)^{\lambda_{ij}^r} - \left( \frac{\sigma_{ij}}{r_{ij}}\right)^{\lambda_{ij}^a} \right], \;\; \textrm{for} \;\; r_{ij} < r_c 
\end{equation}
where
\[ C=\left( \frac{\lambda_{ij}^r}{\lambda_{ij}^r-\lambda_{ij}^a} \right) \left( \frac{\lambda_{ij}^r}{\lambda_{ij}^a}  \right)^{\left( \frac{\lambda_{ij}^a}{\lambda_{ij}^r-\lambda_{ij}^a} \right)}. \]
The indices $i$ and $j$ indicate the bead type (\textit{e.g.}, W, M, \textit{etc.}). Thus,
$\sigma_{ij}$, $\varepsilon_{ij}$, $\lambda_{ij}^r$, and $\lambda_{ij}^a$ are parameters of the
Mie potential, while $r_{ij}$ is the distance between any two beads. The values of the Mie potential
parameters for different pairs of beads are summarised in Table~\ref{table1}; the potential
cutoff is set to $r_c=4.583\sigma$. 
In addition, $\lambda_{ij}^a = 6$, irrespective of the bead type \cite{Ramrattan2015}. 

Chain molecules are built by tethering subsequent effective beads together using a harmonic potential,
\begin{equation}
\label{harmonicbond}
V(r_{ij}) = 0.5 k (r_{ij}-\sigma_{ij})^2,
\end{equation}
where values of $\sigma_{ij}$ are given in Table~\ref{table1},   
and $k=295.33 \varepsilon/\sigma^2$.
$\sigma$ is the unit of length while $\varepsilon$ is the energy unit.
Moreover, EO effective beads along the chain interact via a harmonic angle potential, of the form,
\begin{equation}
\label{harmonicangle}
V_{\theta}(\theta_{ijk}) = 0.5 k_{\theta} (\theta_{ijk}-\theta_0)^2,
\end{equation}
where $\theta_{ijk}$ is the angle between three consecutive beads along a chain, 
$k_{\theta}=4.32 \varepsilon/$rad$^2$ is a constant indicating the strength of the harmonic interaction
(stiffness of the chain), and $\theta_0=2.75$~rad is the 
equilibrium angle.

The fluid--substrate interactions were taken into account by integrating the solid potential 
considering wall composed of spherical Mie beads (implicit substrate) 
resulting in the following expression \cite{Forte2014}:
\begin{equation}
\label{eq:sub}
U_{sub}(D) = 2 \pi\rho C \varepsilon_{ij}\sigma_{ij}^3
   \left[ A  \left( \frac{\sigma_{ij}}{D} \right)^{\lambda_{ij}^r-3} 
   - B \left( \frac{\sigma_{ij}}{D} \right)^{\lambda_{ij}^a-3} \right]. 
\end{equation}
Here, $D$ is the vertical distance between beads and the substrate,
$A=1/(\lambda_{ij}^r-2)(\lambda_{ij}^r-3)$ and $B=1/(\lambda_{ij}^a-2)(\lambda_{ij}^a-3)$.
$C$, $\sigma_{ij}$,$\varepsilon_{ij}$, $\lambda_{ij}^r$, and $\lambda_{ij}^a$ have been defined in Eq.~\ref{Mie},
and $\rho$ is the number density, which typically for a paraffinic substrate is $\rho\approx 1\sigma^{-3}$. 
For the substrate potential the cut-off distance is the same as in the case of fluid--fluid interactions.
In the case of the substrate--water (SW) interaction, the strength of the interaction is chosen 
to provide a contact angle of approximately $60^\circ$. To achieve this, the value of the
parameter $\varepsilon_{ \rm SW}=1.4 \varepsilon$.  The respective values for
the substrate $\sigma_{\rm SS}=\sigma$ and all fluid--solid interactions 
can be obtained by employing common combining rules \cite{Lafitte2013}, namely,
$\sigma_{ij}=(\sigma_{ii}+\sigma_{jj}/2)$, $\lambda^r_{ij} - 3 = \sqrt{(\lambda^r_{ii}-3)(\lambda^r_{jj}-3)} $,
and $\varepsilon_{ij}= (1-k_{ij}) \sqrt{\sigma^3_{ii}\sigma^3_{jj} \varepsilon_{ii} \varepsilon_{jj}} / \sigma^3_{ij} $
\cite{Lafitte2013}. 
Our model has been matched to experimental data at all stages
of the method development and the acquired data were compared with experimental
results, with the coarse-grained model specifically parameterised to reproduce:
the experimental phase behaviour of water and surfactants \cite{Lobanova_PhD,Lobanova2015,Lobanova2016,Theodorakis2015a,Theodorakis2015b}, the 
spreading behaviour \cite{Theodorakis2015a,Theodorakis2015b}, and observed
effects of surfactant architecture and bilayer formation \cite{Hill1998,Ruckenstein1996}.

All simulations were performed in the NVT ensemble
by using the Nos\'e-Hoover thermostat as implemented in the HOOMD package \cite{Anderson2008} 
based on the MKT equations \cite{Martyna1994,Cao1996}, with an integration time-step of
$\Delta t = 0.005 \tau$, where $\tau = \sigma (m/ \varepsilon)^{1/2}$ is the time unit.
The reduced time unit corresponds to $1.4062$~ps. While the size of the simulation 
box and the number of particles remain constant during the simulation, the temperature fluctuates around a predetermined
value, which in our case is $T = 0.6057$ (corresponding to $25^{\circ}C$). The simulation box is $201\sigma$ long in
the $x$ and $y$ directions guaranteeing that periodic images of the droplet do not interact with each other
due to the presence of periodic boundary conditions in these directions. We have also placed two walls in 
the $z$ (normal) direction. The bottom wall is implicit and expressed through Eq.~\ref{eq:sub} that
corresponds to an unstructured wall of infinite thickness. The top wall is represented
by LJ beads that interact with the rest of the system through a purely repulsive LJ potential.
We typically place $8\times10^4$ beads
in the simulation box and bring the droplet in an equilibrium position.
as follows: Water and surfactant beads are placed close to each other 
(\textit{e.g.} in an FCC structure as is usually done in MD simulation)
and, also,
close to the substrate. Then, running the MD simulation will
lead to the formation of a spherical droplet, which can attach to
the substrate. At this stage, the potential between the surfactants
and the substrate is switched off. Hence, this initial configuration
roughly corresponds to the equilibrium state 
of the aqueous droplet on the paraffinic substrate. 
Once the droplet reaches the equilibrium, we switch on the 
potential between the surfactants and the substrate and the nonequilibrium spreading process starts until
a new equilibrium state is reached by the system. Typical trajectories to reach the latter equilibrium depend
on the type of surfactant and its concentration. Here, we have considered the same concentration for all cases
in order to enable the comparison between the different cases of superspreading and nonsuperspreading processes.
This concentration is in the superspreading regime and is 6.3 $\times$ Critical Aggregation Concentration
of Silwet-L77 \cite{Theodorakis2015a}.
In general, trajectory lengths were in the range $10^7$ to $10^8$ MD steps. Trajectory
samples were gathered every $10^4$ MD steps for all cases, which ensures an independent statistical collection
of snapshots required for our analysis while guaranteeing the acquisition of
adequate number of snapshots for describing nonequilibrium spreading processes.

\section*{Superspreading}
\begin{figure}
\includegraphics[scale=0.58]{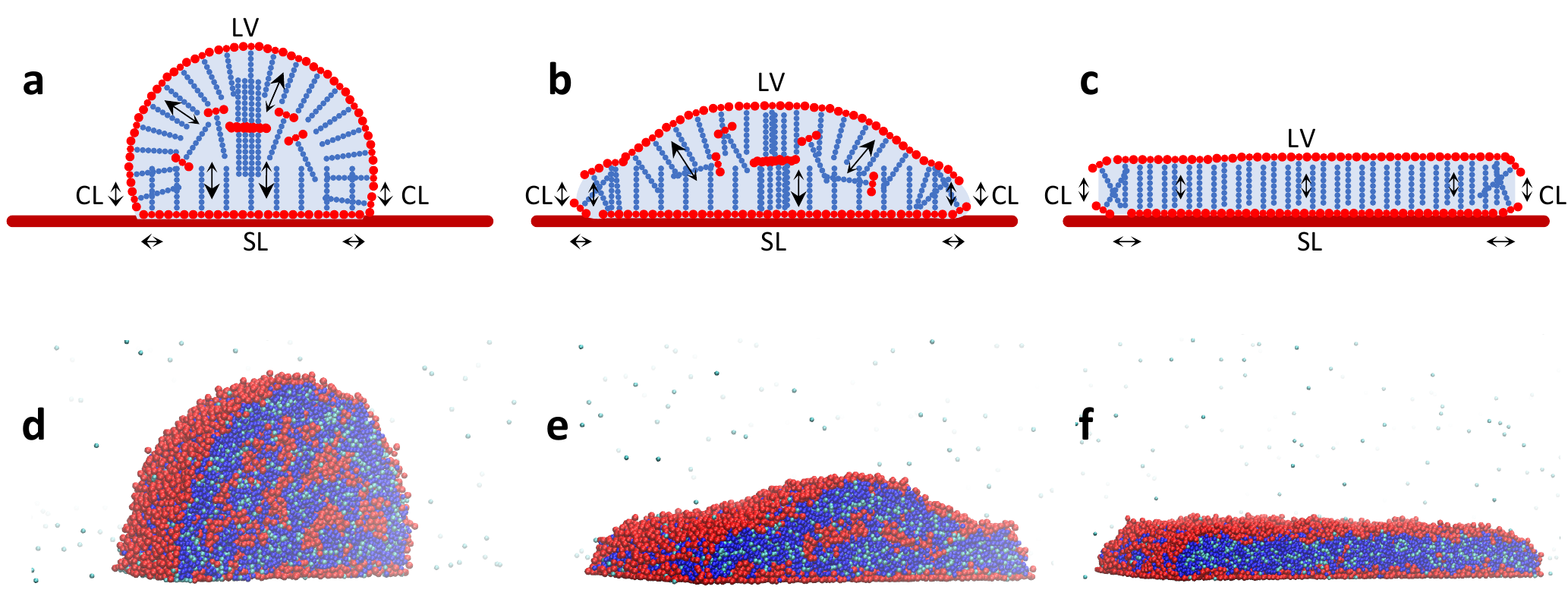}
 \caption{\label{fig1}Schematic drawings (a-c) illustrating the dominant directions of adsorption processes during
 the superspreading of an aqueous droplet laden with Silwet-L77 surfactants and representative snapshots (d-e).
 `LV' indicates the liquid--vapour interface, `SL' the solid--liquid interface, and `CL' the contact line.
 In the snapshots, red indicates the hydrophobic groups of the surfactants, while hydrophilic groups are
 in blue. Water molecules are in cyan. The cross-section of each droplet
 is shown in order to clearer illustrate the distribution of surfactant
 within the droplet.}
\end{figure}

In previous work \cite{Theodorakis2015b}, we used CG MD simulation 
to propose a mechanism for superspreading of surfactant-laden aqueous droplets, 
based on a detailed molecular-level analysis of adsorption processes that take
place in the droplet during spreading. 
Figure~\ref{fig1} schematically illustrates these processes at three different stages of spreading in the case of an aqueous droplet with superspreading surfactant (Silwet-L77), 
namely, an initial and an intermediate nonequilibrium
stage (Fig.~\ref{fig1}a, b respectively) and an equilibrium final stage (Fig.~\ref{fig1}c) 
of a bilayer conformation.
The spherical-cap shape of the droplet gradually transforms into the bilayer structure
through the expansion of the contact line as shown in Fig.~\ref{fig1}. During this transformation, 
the directions of surfactant adsorption/desorption processes are illustrated by arrows (Fig.~\ref{fig1}a, b, c). 
While a number of different adsorption processes take place during superspreading, 
two of them are essential to sustain the
rapid spreading of the droplet. The first one is the adsorption of surfactant from 
the liquid--vapour (LV) interface onto the solid--liquid (SL) interface (substrate)
at the contact line (CL), which has been previously suggested by continuum modelling \cite{Karapetsas2011}.
This adsorption of surfactant at the CL results in an increase in the area
of SL and LV interfaces, which causes a temporary depletion of surfactant
at the interfaces. For this reason, the fast repletion of the interfaces with surfactant from the
bulk is crucial in order to sustain the rapid spreading of the droplet. As a result, the ability
of effectively adsorbing at the interfaces is an essential feature of superspreading surfactants
 \cite{Theodorakis2015b}. In the case of the bilayer structure at the final stages of spreading, 
surfactants between different parts of the droplet continuously exchange. In this case, there
is no dominant direction of adsorption and the system is in a dynamic equilibrium.
Characteristic snapshots of aqueous
droplets laden with Silwet-L77 surfactants as obtained by MD simulation are presented in Fig.~\ref{fig1}, 
where the distribution of surfactant molecules within the droplet and at the interfaces is illustrated. 
As the droplet spreads towards the final equilibrium bilayer structure, the aggregates from
the bulk supply the interfaces with surfactant. In the final stage, aggregates have dissolved
and only surfactant monomers appear within the bilayer.

\section*{Results and Discussion}

\begin{figure}
\centering
\includegraphics[scale=0.40]{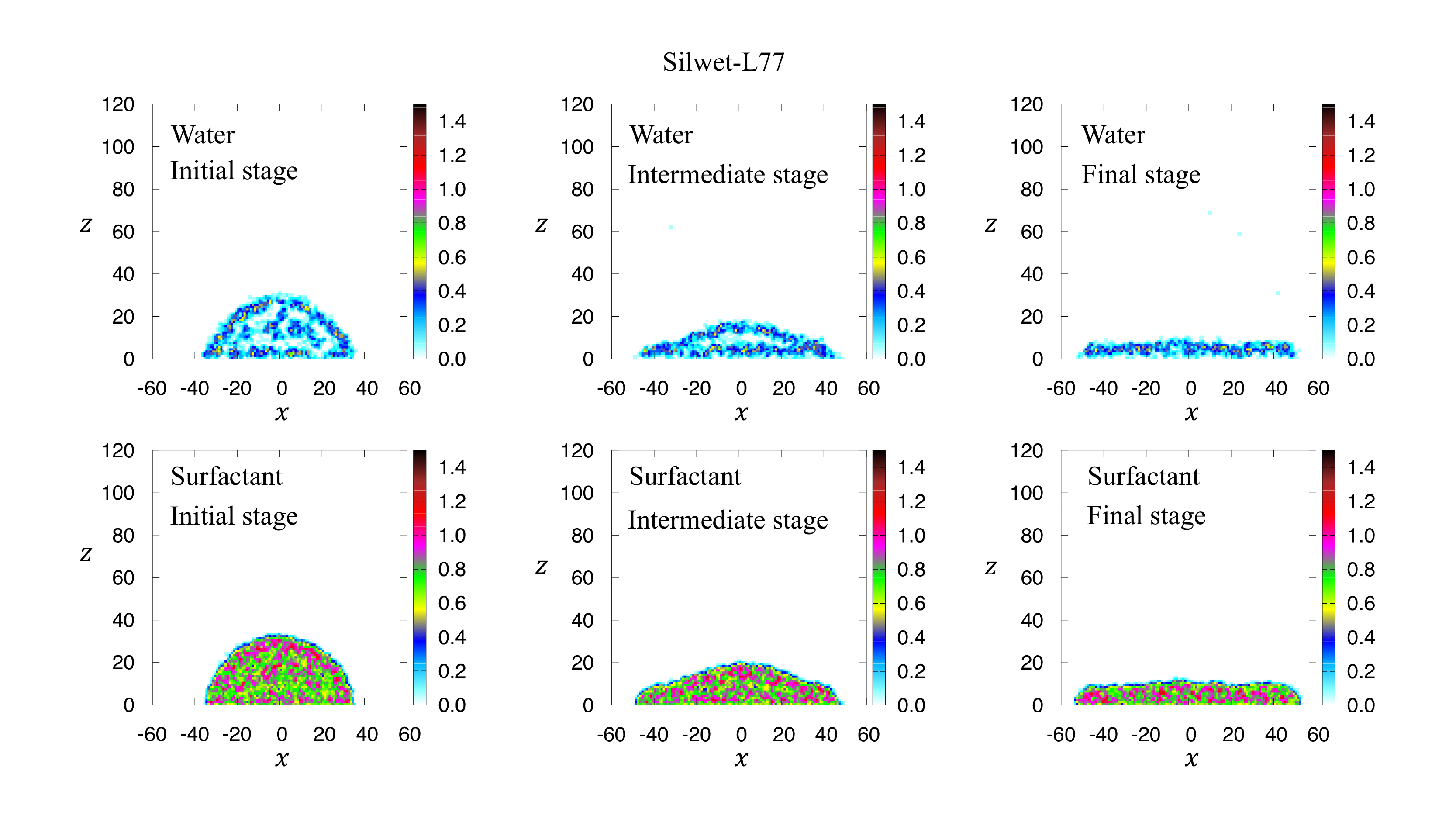}
 \caption{\label{fig2} Density profiles of water and  surfactant molecules (Silwet-L77) at a cross-section
 of the droplet (perpendicular to the substrate)
 at an initial, intermediate, and final stage of superspreading, as indicated.
 The different colours indicate the density. }
\end{figure}

A quantitative assessment of the distribution of water and surfactant molecules within the droplet at different times can be obtained by calculating the density profiles at the cross-section. 
The cross-section is perpendicular to the substrate and passes through the centre
of mass of the droplet. In the case of droplets with
Silwet-L77 (superspreading surfactant, Fig.~\ref{fig2}),
the density profiles of water molecules indicate
the absence of water molecules at certain areas inside the droplet,
where the hydrophobic cores of surfactant aggregates are present. 
On the contrary, water molecules are homogeneously
distributed along the LV and the SL interfaces, which are initially 
connected with the water domains within the droplet. As the droplet spreads, 
the SL and LV interfaces merge at the CL as is illustrated by the water profiles in Fig.~\ref{fig2}, which results in the formation of a bilayer. At an intermediate stage of the
spreading process, the bulk of the droplet is dominated by the hydrophobic
cores of the surfactant, while at the final stage water molecules are distributed
homogeneously along the bilayer and between the surfactants at the SL and LV
interfaces. During the spreading process, the fluctuation in the number
of water molecules in the droplet is very small (below 1\%) and a slight increase in water
is observed as surfactants occupy the interfaces of the droplet during the bilayer formation.
The density profile of surfactant molecules 
indicates a distribution of surfactant without large deviations from the
average density at all stages of 
spreading. Finally, the density distribution of surfactant
indicates the boundaries of the droplet as the hydrophobic parts are exposed to the air
at the LV interface. The spreading of the droplet is symmetric in all directions onto
the $x-y$-plane as a result of the smooth unstructured substrate.

\begin{figure}
\centering
\includegraphics[scale=0.34]{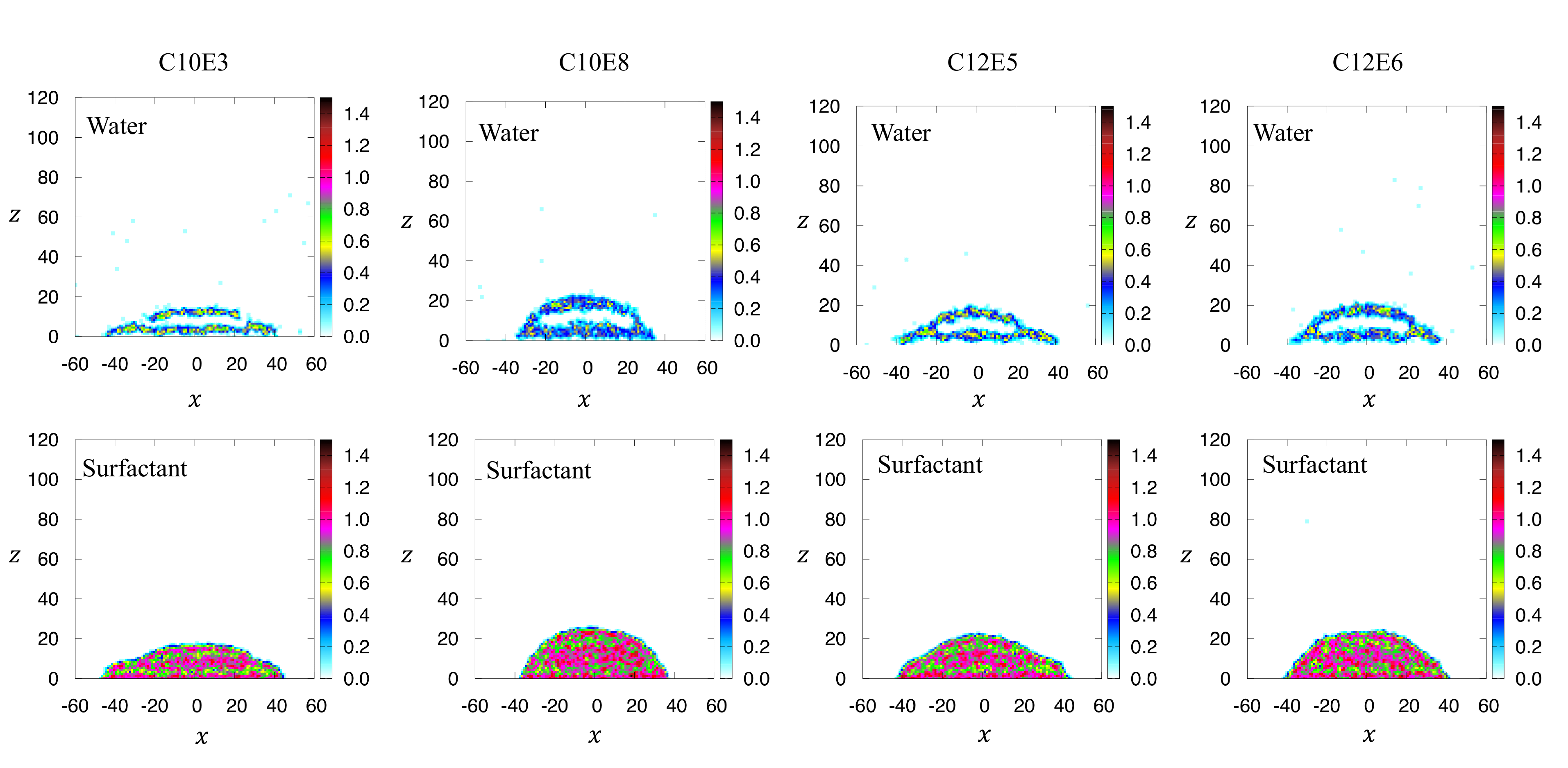}
 \caption{\label{fig3} Density profiles of water and  surfactant molecules  at a cross-section
 of the droplet (perpendicular to the substrate) for various common surfactants, as indicated. The different colour indicates the density. For the cases of C10E3, C10E8, and C12E6, 
 a representative snapshot of the equilibrium state is shown, whereas in the case of C12E5
 a representative snapshot of an intermediate spreading stage is illustrated.}
\end{figure}

We now compare these profiles to typical profiles for common surfactants (Fig.~\ref{fig3}).
For the range of common surfactants considered in this study, only droplets with C12E5 surfactant eventually form a bilayer structure, but the measured spreading exponent for this
case is below the range for superspreading behaviour (0.16--1) at about 0.1. 
In fact, the density profile of water in the case of C12E5 at an intermediate stage of
the spreading process shares some similarities with the case of superspreading due to the
initial formation of the bilayer. Although the formation of a bilayer always takes place
in the case of superspreading (\textit{e.g.} Silwet-L77, T3E3), 
certain common surfactants (\textit{e.g.} C12E5) can lead to the formation
of bilayer as well, as demonstrated by our computer simulations. Hence, the bilayer
structure is characteristic of superspreading behaviour, but it is not a necessary condition.
A careful comparison of the density profiles between superspreading and common
surfactants indicates a larger heterogeneity of surfactant distribution within the droplet
in the case of common surfactants,
which reveals an aggregation preference that prevents the surfactants from
leaving these aggregates and adsorbing at the interfaces. Indeed, in droplets with common surfactants (Fig.~\ref{fig3})
the hydrophobic cores of surfactant aggregates can be better distinguished than in droplets with superspreading surfactant, since local density variations from
the average density are larger.
Below, we will discuss this point in more detail.

\begin{figure}
\centering
\includegraphics[scale=0.60]{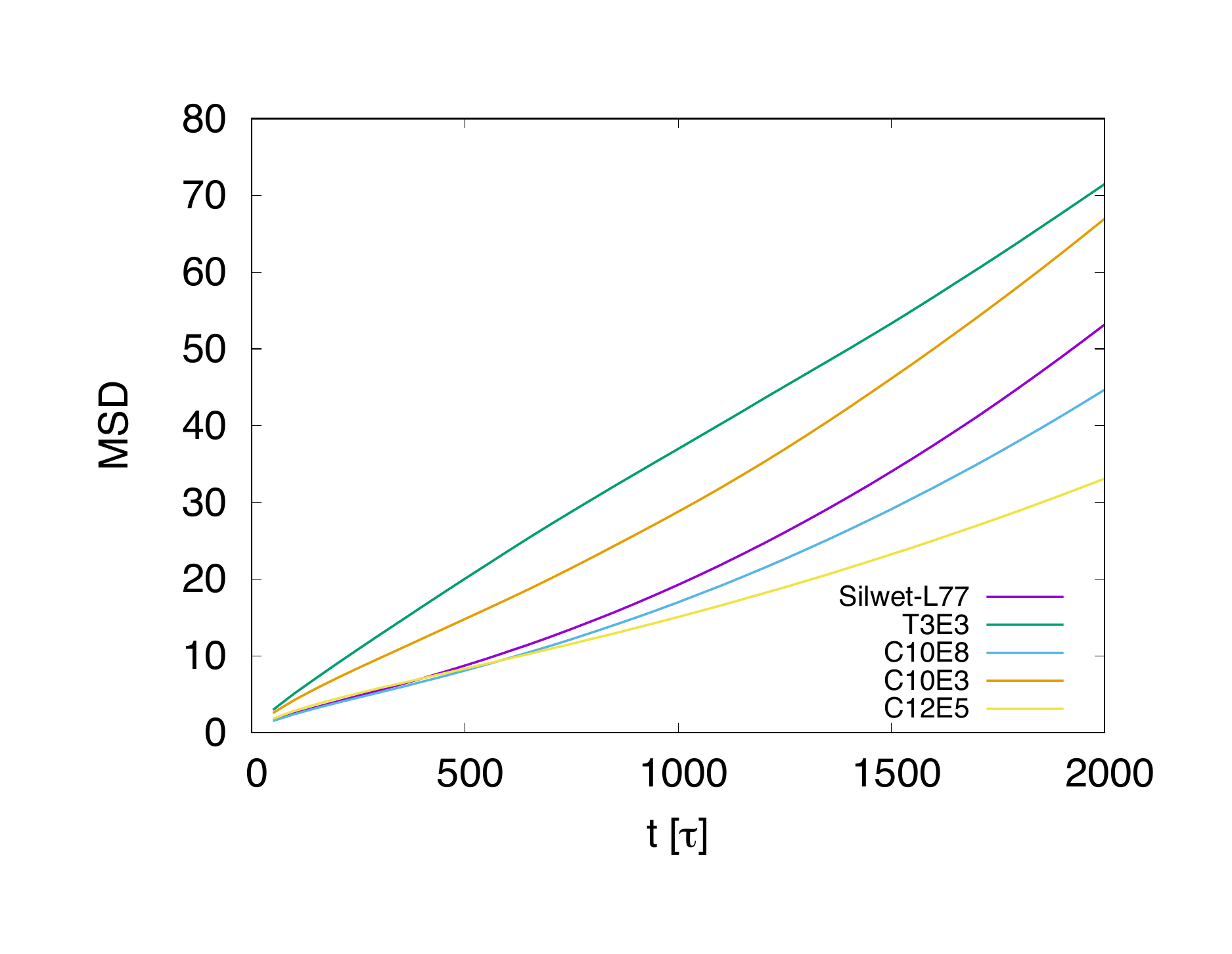}
 \caption{\label{fig4} Mean-square displacement for different surfactant
 as indicated.}
\end{figure}

Molecular dynamics simulation can provide estimates of the adsorption rates of surfactant at different parts of the droplet
by tracking individual chains, which is an important strength of molecular-level
simulation. 
Considering an average of the adsorption rates in the ensemble of 
all surfactants by analysing the individual trajectories of each surfactant
during the spreading of the droplet we have confirmed the importance of
the surfactant adsorption at the LV interface as well as at the SL
interface and to a lesser extent to the CL, and
from the bulk to the SL and LV interfaces. These adsorption processes are crucial for the
formation of bilayer, but spreading rates vary depending on the ability of the surfactant
to replenish the LV and SL interfaces, where a temporary depletion of surfactant is
observed during droplet spreading. Our conclusions have been corroborated by the density
profiles, which show a tendency of surfactants to remain in aggregates
for droplets with common surfactants. 
Indeed, by following the trajectory of each surfactant molecule within
the droplet during spreading we can measure the probability of finding surfactants in
different parts of the droplet (Table~\ref{table2}).
We have confirmed that common
surfactants have a higher tendency to remain in the bulk in the form of aggregates, 
as the density profiles of Fig.~\ref{fig3} might suggest. Our results also indicate that 
common surfactants exhibit smaller or comparable probabilities of being at
the LV interface and smaller probability of being at the SL interface due to the stronger
hydrophobic character of the lyophilic part of the surfactant molecules (cf. interactions
in Table~\ref{table1}).
This also affects the ability of surfactant to adsorb at the CL.
Finally, among the common surfactants considered here, the droplet with C12E5 surfactants 
is able to form the bilayer structure. In this case, we found
the smallest probability that surfactants will stay in the bulk of the droplet. 
Analysis of the overall diffusion of surfactant molecules 
within the droplets (Fig.~\ref{fig4}) has further underlined the importance of adsorption/desorption processes and the aggregation tendency of surfactants 
in the spreading process. The overall motion of surfactant within the droplet
is subdiffusive irrespective of whether the surfactant is a superspreader. 
While differences between cases are small, our results suggest that both the size
(smaller molecules generally tend to diffuse faster), chemistry
and molecular architecture may affect the motion of surfactant in the droplets, but
differences are small. Based on this comparison between common and superspreading surfactants
we have concluded that the molecular diffusion plays a minor role
in the spreading process and the ability of surfactants to replenish the interfaces
during spreading is mainly dictated by the aggregation properties of that surfactant.

\begin{figure}
\centering
\includegraphics[scale=0.65]{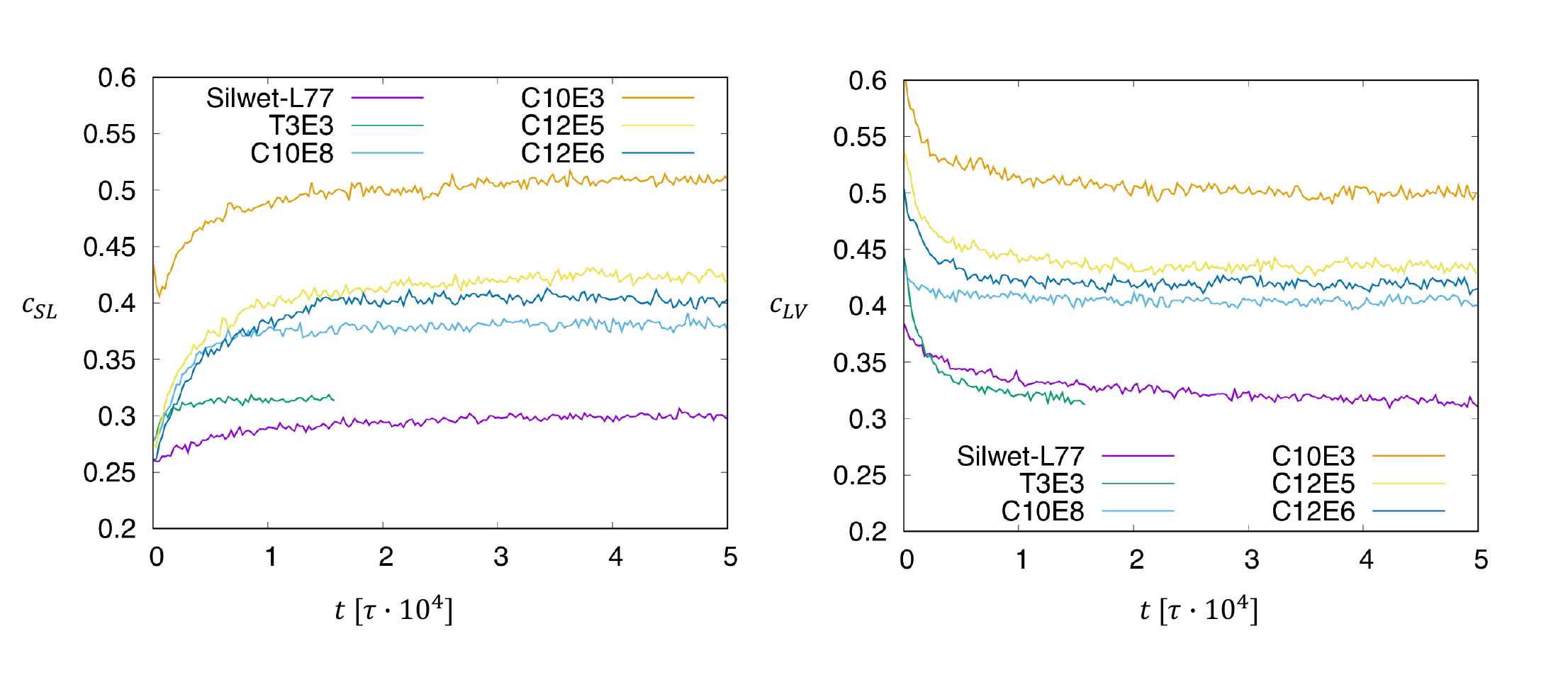}
 \caption{\label{fig5} The time evolution of the number of surfactant molecules per area at the SL interface (left panel), $c_{SL}$, 
 and the LV surface (right panel), $c_{LV}$.}
\end{figure}

\begin{figure}
\centering
\includegraphics[scale=0.35]{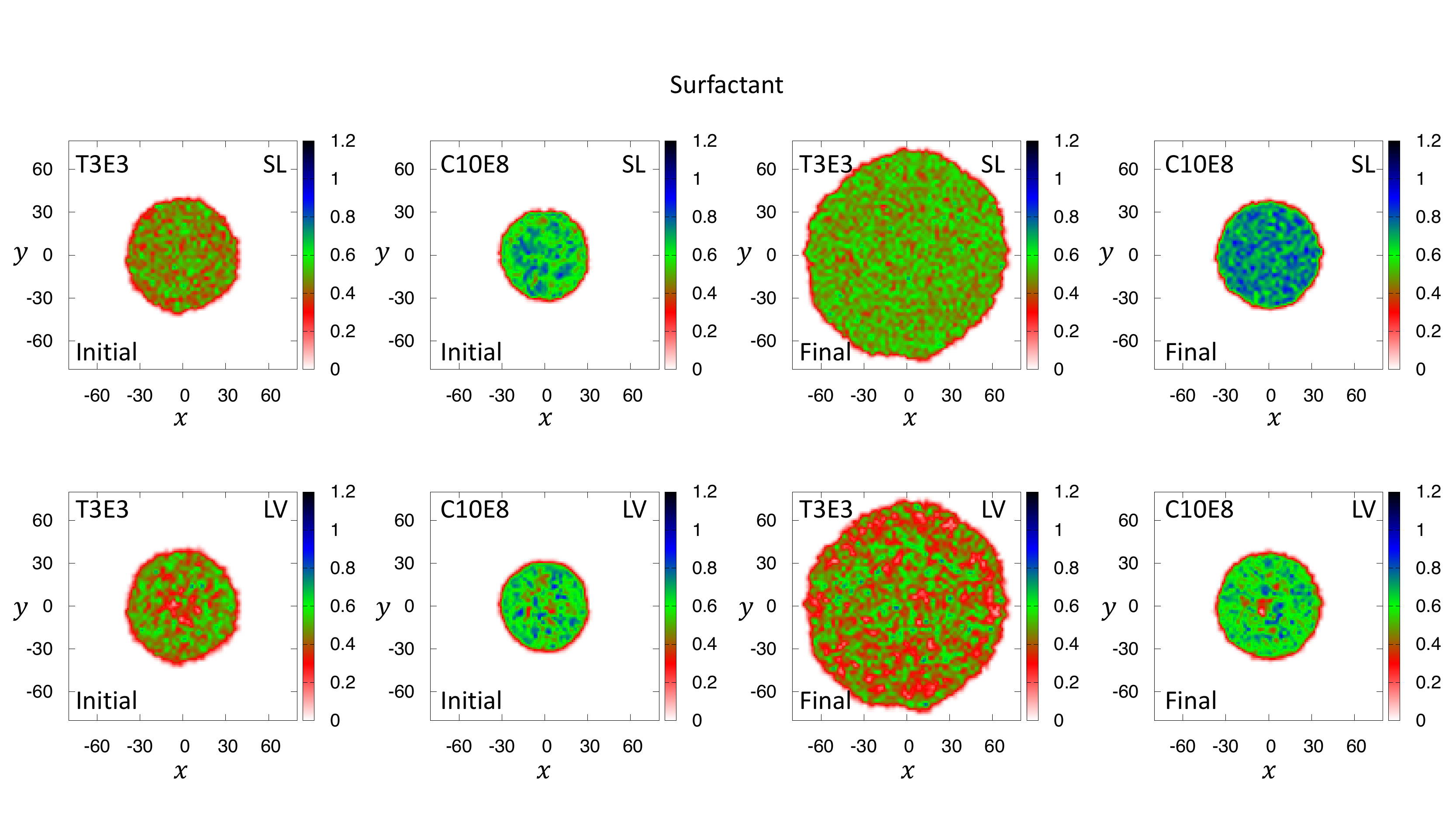}
 \caption{\label{fig6} Density profiles of surfactant molecules
 at the SL interface and the LV surface for two different surfactant
 cases at an initial and a final stage of spreading as indicated. 
 The different colour indicates the density on the interfaces.}
\end{figure}

\begin{figure}
\centering
\includegraphics[scale=0.35]{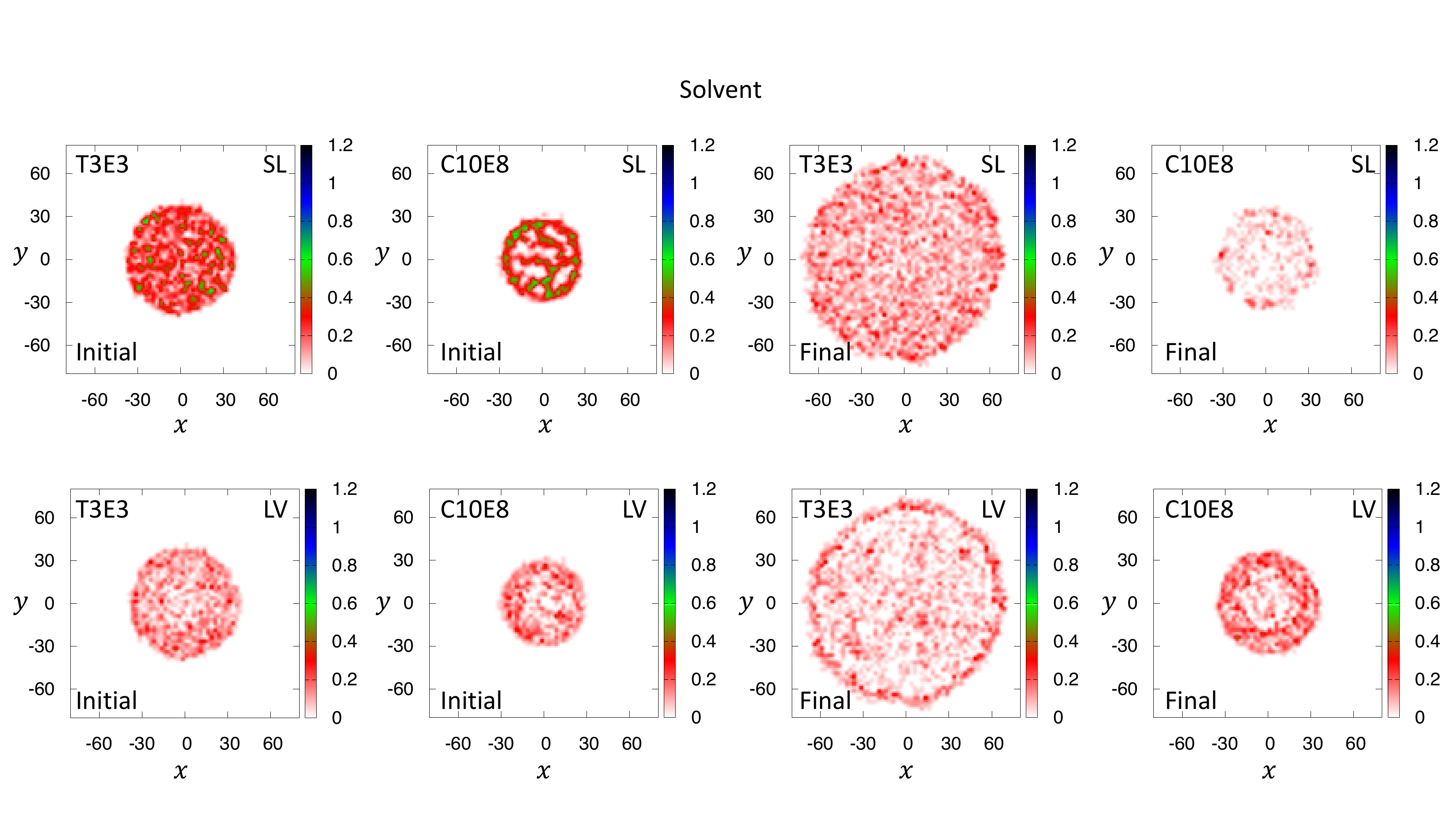}
 \caption{\label{fig7} Similar to Fig.~\ref{fig8}, but the density
 of solvent molecules (water) is shown.}
\end{figure}

Fig.~\ref{fig5} illustrates the time evolution of the number of surfactant molecules per area at the
SL interface, $c_{SL}$,  and the LV surface, $c_{LV}$. $c_{SL}$ is increasing at the initial stages of spreading as
the interaction between surfactant molecules and the substrates is switched on reaching
a constant value when the final dynamic equilibrium is established. 
In contrast, $c_{LV}$ decreases at the initial steps and remains 
constant during the spreading process. In the case of superspreading surfactants, 
small changes are observed in the values of  $c_{LV}$ and $c_{SL}$. At the final equilibrium
bilayer stage, the values of $c_{LV}$ and $c_{SL}$ remain constant (not shown here).
In the case of the LV surface, there is not
an additional interaction (\textit{e.g.} the substrate potential) to attract the molecules
at the LV surface. As a result, the spreading process leads to an initial depletion
of surfactant at the LV surface and a constant supply of surfactant beyond this 
initial stage. Overall, the values of $c_{LV}$ and $c_{SL}$ depend on the
molecular architecture. For example, linear surfactants allow for a better packing at the interfaces.
In other words, a larger number of molecules are required to cover the interfaces. 
This is also illustrated as we attempt to compare the density profiles at the SL and LV interfaces
of droplets with different surfactants. 
Fig.~\ref{fig6} illustrates a typical comparison between superspreading (T3E3) and common (C10E8) spreading
cases at the initial and final stages of the spreading process. The density of surfactants
appears larger in the case of the linear C10E8 surfactant than in the case of the T-shaped T3E3 surfactant, despite 
the overall considerably smaller size of the latter. While this difference in behaviour 
is larger at the SL interface, a smaller difference is observed at the LV surface. 
Considering the density profiles at the LV and SL interfaces for
the solvent molecules (Fig.~\ref{fig7}), we found that the interfaces of the droplets
are dominated by the surfactants and the solvent's density is rather small. The 
presence of water at the interfaces is smaller in the case of the C10E8 surfactant
due to the closer packing of the molecules. At the LV interface the T3E3 has formed
a bilayer, which is dominated at the CL by water, while the rest of the interface is
dominated by surfactant at the bilayer. 

\begin{figure}
\centering
\includegraphics[scale=0.60]{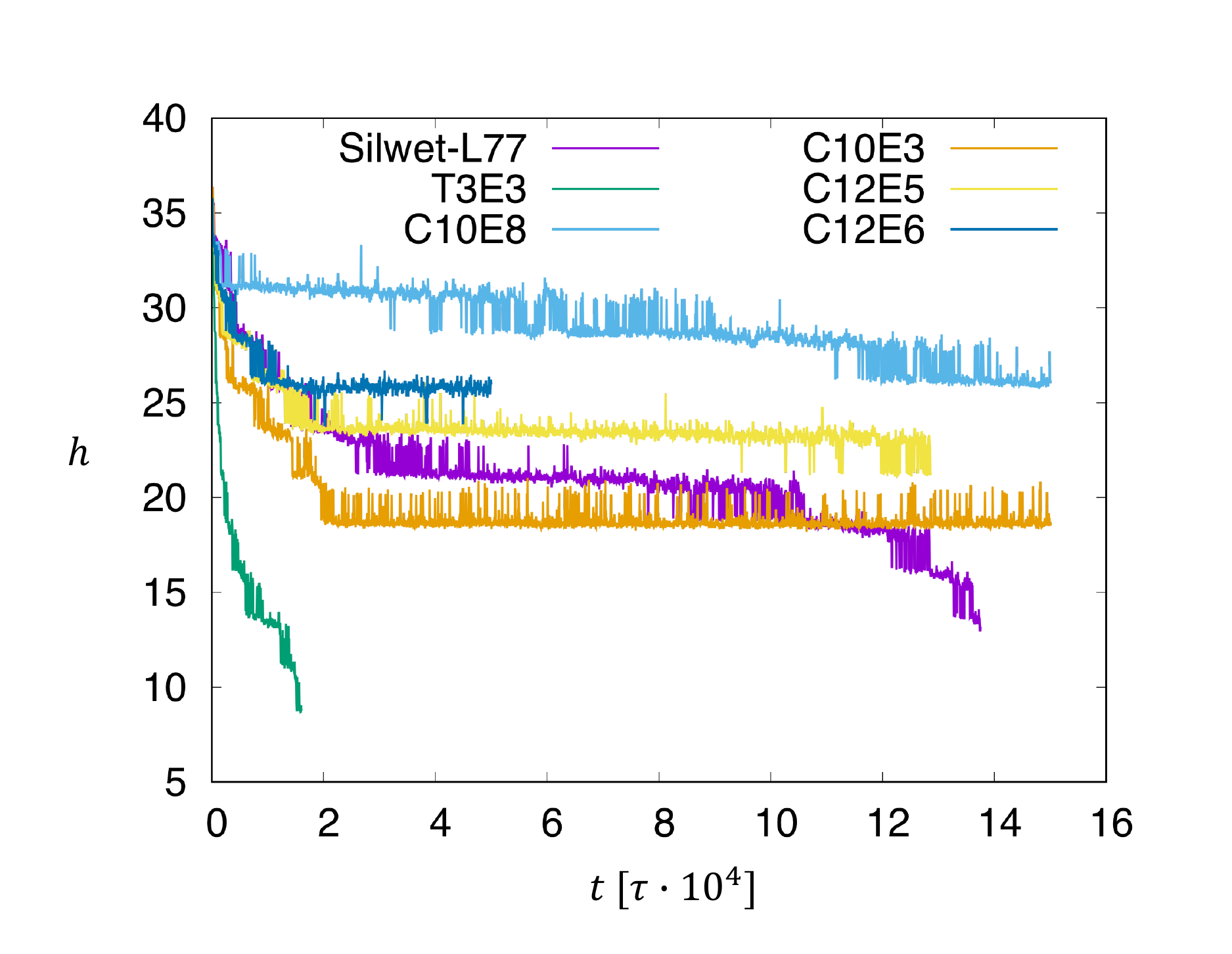}
 \caption{\label{fig8} Time evolution of droplet height 
 (measured from the apex to the SL
 interface) for droplets with different surfactant as indicated.}
\end{figure}

\begin{figure}
\centering
\includegraphics[scale=0.60]{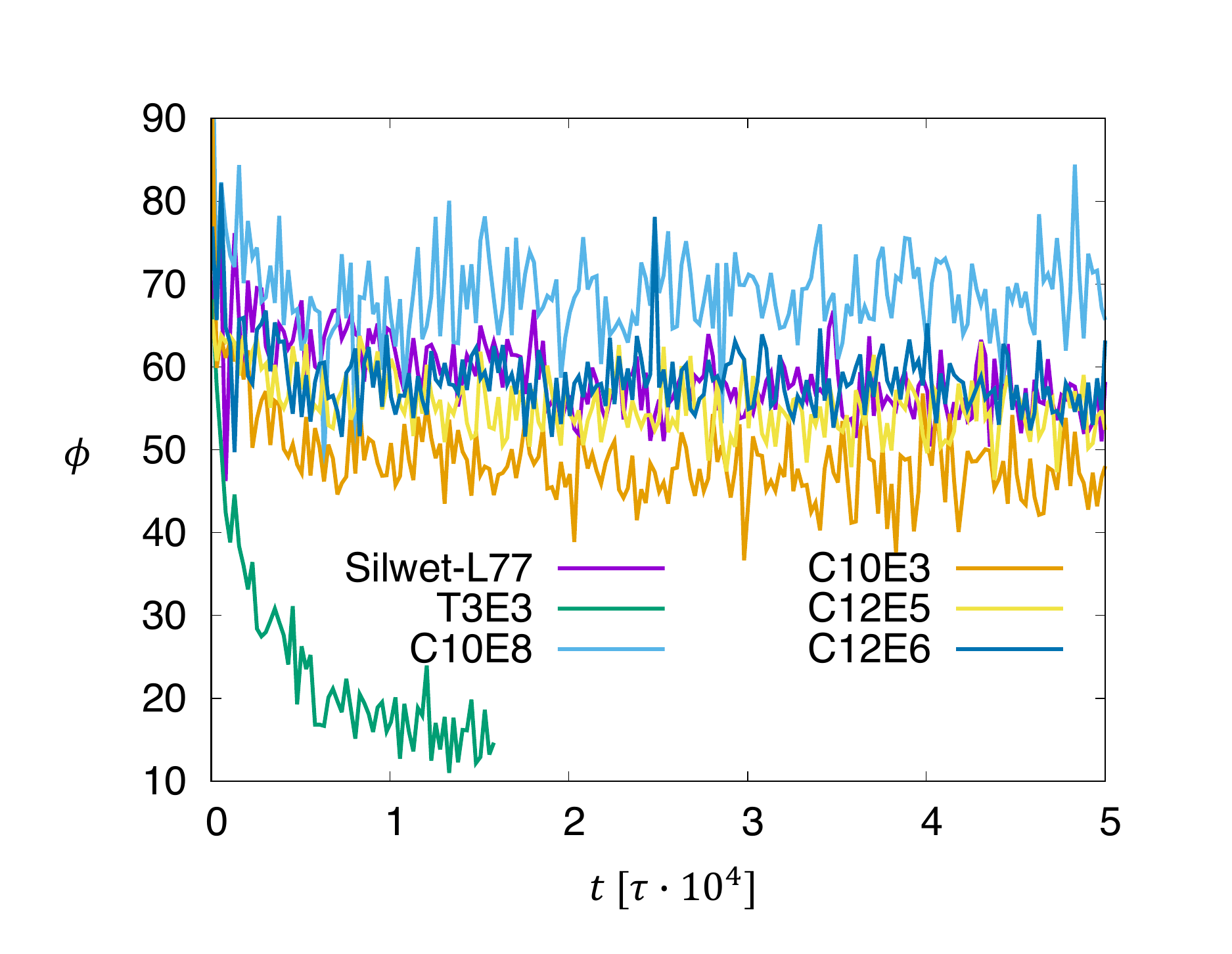}
 \caption{\label{fig9} Contact angle $\phi$ for droplet with 
 different surfactant as indicated.}
\end{figure}

Fig.~\ref{fig8} illustrates the time dependence of the droplet height for different surfactant cases.
The droplet height is measured from the droplet apex to the SL interface. Hence, there is no
direct information about the formation of bilayer when measuring the droplet height.
However, we know that the formation of bilayer takes place in the case of droplets with Silwet-L77, T3E3, and C12E5
surfactants. The fluctuations in the height for all cases is a consequence of the 
continuous depletion and refill of the LV interface with surfactant from the bulk. 
Droplets with surfactants of smaller size tend to have a smaller droplet height.
A comparison between the Silwet-L77 and T3E3 also indicates that the height of
the bilayer in the case of T3E3 is smaller than in the case of Silwet-L77. The 
bilayer height of C12E5 surfactant is larger than in the case of Silwet-L77. This
is also reflected in the case of the contact angle (Fig.~\ref{fig9}), where smaller surfactants lead
to smaller contact angles. In order to measure the contact angle, we have used a
linear fit for the LV interface at the CL. Clearly, the formation of bilayer at the
CL does not allow for a strict interpretation of our measurements and a more accurate
way of measuring the contact angle at the CL, based on the droplet curvature
\cite{Theodorakis2015a}, is clearly not applicable in this case.
Moreover, we do not attempt here to describe the 
fluctuating contact angle behaviour, which has been considered
in a previous study in detail \cite{Smith2016}.
However, our measurements of the contact angle at the CL 
are consistent with the fact 
that smaller surfactants are associated with smaller contact angles.

\section*{Conclusions}
In this study, we have discussed various properties of droplets laden with common and superspreading
surfactants. Our analysis has confirmed that the ability of surfactant to adsorb at the interfaces
is a key feature of the superspreading mechanism. We have also found that a key feature of nonsuperspreading
surfactants is their higher tendency towards aggregates formation, which is a result of the stronger
lyophilic interactions. Moreover, the surfactant spatial distribution,
diffusion and chain architecture play a smaller role in the spreading process. Finally, we have discussed various properties of the droplets, including height, contact angle and surface concentration, in the context of superspreading behaviour of the droplets.
These highlight the different behaviour of common and superspreading surfactants. 
We anticipate that our work will provide further insight into the spreading mechanisms of surfactant-laded droplets and will underline differences between common and superspreading surfactants.
This insight will directly benefit applications requiring the control of wetting through a rational chemical and architectural design of surfactants.

\section*{Acknowledgements}
The authors would like to thank Omar Matar and Richard Craster for
valuable discussions and support on the topic of this study from its very initial
stage.
This project has received funding from the European Union's Horizon 2020 research and innovation programme under the 
Marie Sk{\l}odowska-Curie grant agreement No.\ 778104.  This research was supported in part by PLGrid Infrastructure.

\singlespacing

\section*{References}

\bibliography{elsarticle-template}

\newpage

\begin{table}
\caption{\label{T2} Reduced molecular parameters of the Mie interaction potential 
between effective beads. $\lambda_{ij}^{a}=6$ for all cases. The length and the energy unit are $\sigma=0.43635$~nm
and $\varepsilon/k_B=492$~K, respectively. Therefore, $k_BT/ \varepsilon = 0.6057$,
which corresponds to 25$^0$C.}
\label{table1}
\begin{tabular}{|c|c|c|c|}
\hline i--j   & $\sigma_{ij} [\sigma]$  & $\varepsilon_{ij} [\varepsilon/k_B]$  & $\lambda_{ij}^r$  \\ 
\hline \multicolumn{1}{|l|}{W--W}   & 0.8584  & 0.8129  & \multicolumn{1}{|l|}{8.00} \\ 
 \multicolumn{1}{|l|}{W--M}   & 1.0491  & 0.8132  & 13.72 \\ 
 \multicolumn{1}{|l|}{W--D}   & 0.9643  & 0.6311  & 10.38 \\ 
 \multicolumn{1}{|l|}{W--EO}  & 0.8946  & 0.9756  & 11.94 \\ 
 \multicolumn{1}{|l|}{W--CM}  & 0.9292  & 0.5081  & 10.75 \\ 
 \multicolumn{1}{|l|}{M--M}   & 1.2398  & 0.8998  & 26.00 \\ 
 \multicolumn{1}{|l|}{M--D}   & 1.1550  & 0.7114  & 18.83 \\ 
 \multicolumn{1}{|l|}{M--EO}  & 1.0853  & 0.8262  & 22.18 \\ 
 \multicolumn{1}{|l|}{M--CM}  & 1.1199  & 0.7800  & 19.61 \\ 
 \multicolumn{1}{|l|}{D--D}   & 1.0702  & 0.5081  & 13.90 \\ 
 \multicolumn{1}{|l|}{D--EO}  & 1.0004  & 0.6355  & 16.21 \\ 
 \multicolumn{1}{|l|}{D--CM}  & 1.0351  & 0.5953  & 14.43 \\ 
 \multicolumn{1}{|l|}{EO--EO} & 0.9307  & 0.8067  & 19.00 \\ 
 \multicolumn{1}{|l|}{EO--CM} & 0.9653  & 0.7154  & 16.86 \\ 
 \multicolumn{1}{|l|}{CM--CM} & 1.0000  & 0.7000  & 15.00 \\ 
\hline 
\end{tabular} 
\end{table}

\begin{table}
    \caption{Probability of surfactant molecules being at different parts of the droplet 
    (SL, LV, CL, BULK) for different surfactants (T3E3, Silwet-L77, C10E3, C10E8, C12E5, C12E6) as indicated. 
    The average over all surfactants
    during the spreading process is considered for calculating this
    probability. The values
    of the probabilities are truncated to the third digit and probabilities add to unity. }
    \label{table2}
    \centering
\begin{tabular}{ |p{1.66cm}||p{1.0cm}|p{1.0cm}|p{1.0cm}|p{1.0cm}|  }
% \hline
% \multicolumn{4}{|c|}{Country List} \\
 \hline
  & SL & LV  & CL  & BULK \\
 \hline
 T3E3        & 0.372  & 0.450   & 0.080  & 0.257  \\
 Silwet-L77  & 0.377  & 0.481   & 0.083  & 0.224  \\
 C10E3       & 0.345  & 0.409   & 0.059  & 0.304  \\
 C10E8       & 0.287  & 0.430   & 0.074  & 0.355  \\
 C12E5       & 0.332  & 0.454   & 0.072  & 0.285  \\
 C12E6       & 0.323  & 0.456   & 0.077  & 0.297  \\
 \hline
\end{tabular}
\end{table}

%Figure captions appear in the end and figure files are sent separately

%\begin{table}
%    \caption{Adsorption rates for aqueous droplets with different surfactant as
%    indicated. The arrow indicates the dominant direction of the adsorption process.}
%    \label{table3}
%    \centering
%\begin{tabular}{ |p{1.66cm}||p{2.1cm}|p{2.1cm}|p{2.1cm}|p{2.1cm}|  }
% \hline
% \multicolumn{4}{|c|}{Country List} \\
% \hline
%  & BULK $\rightarrow$ LV  & BULK $\rightarrow$ SL  & LV $\rightarrow$ CL  & SL $\rightarrow$ CL \\
% \hline
% T3E3        & 0.0015571  & 0.0012415  & 0.0001085  & 0.00010430  \\
% Silwet-L77  & 0.0001804  & 0.0078830  & 0.0000180  & 0.00000037 \\
% C10E3       & 0.0001317  & 0.0001003  & 0.0000115  & 0.00000031 \\
% C10E8       & 0.0001424  & 0.0000643  & 0.0000244  & -0.0000141 \\
% C12E5       & 0.0004218  & 0.0003002  & 0.0000243  & 0.00000984 \\
% C12E6       & 0.0004054 & 0.00027062  & 0.0000340  & 0.00000110 \\
% \hline
%\end{tabular}
%\end{table}

\end{document}